\documentclass[a4paper,fleqn,usenatbib]{mnras}

\usepackage{newtxtext,newtxmath}

\usepackage[T1]{fontenc}
\usepackage{ae,aecompl}

\usepackage{graphicx}	
\usepackage{longtable}

\title[Ultraviolet Luminous Objects in GALEX data]
{Search for Ultraviolet Luminous Objects in GALEX data}

\author[S. V. Karpov]{
Sergey V. Karpov,$^{1,2,3}$\thanks{E-mail: karpov.sv@gmail.com}\\
$^{1}$CEICO, Institute of Physics, Czech Acad. Sci., 182 21 Prague 8, Czech Republic \\
$^{2}$Special Astrophysical Observatory, Russian Acad. Sci., Nizhnij Arkhyz 369167, Russia \\
$^{3}$Kazan Federal University, Kazan 420008, Russia \\
}

\author[S. V. Karpov, O. Yu. Malkov and G. Zhao]{
Sergey V. Karpov,$^{1,2,3}$\thanks{E-mail: karpov.sv@gmail.com}
Oleg Yu. Malkov$^{4}$ and Gang Zhao$^{5}$
\\
$^{1}$Institute of Physics, Czech Acad. Sci., 182 21 Prague 8, Czech Republic \\
$^{2}$Special Astrophysical Observatory, Russian Acad. Sci., Nizhnij Arkhyz 369167, Russia \\
$^{3}$Kazan Federal University, Kazan 420008, Russia \\
$^{4}$Institute of Astronomy, 48 Pyatnitskaya St., Moscow 119017, Russia \\
$^{5}$Key Laboratory of Optical Astronomy, National Astronomical Observatories, Chinese Academy of Sciences, Beijing 100012, China
}

\date{Accepted XXX. Received YYY; in original form ZZZ}

\pubyear{2020}

\begin{document}
\label{firstpage}
\pagerange{\pageref{firstpage}--\pageref{lastpage}}
\maketitle

\begin{abstract}
Selection of extreme objects in the data from large-scale sky surveys is a powerful tool for the detection of new classes of astrophysical objects or rare stages of their evolution. The cross-matching of catalogues and analysis of the color indices of their objects is a usual approach for this problem which has already provided a lot of interesting results. However, the analysis of objects that are found in only one of the surveys, and absent in all others, should also attract close attention, as it may lead to the discovery of both transients and objects with extreme color values.
Here we report on the initial study aimed at the detection of objects with a significant UV excess in their spectra by cross-matching of the GALEX all-sky catalogue with several other surveys in different wavelength ranges and analyzing the ones visible in GALEX only, or having extreme UV to optical colors (ultraviolet luminous objects).
We describe the methodology for such investigation, explain the selection of surveys for this study, and show the initial results based on the search in a small fraction of the sky. %
We uncovered several prominent UV-only objects lacking the counterparts in the catalogues of longer wavelengths, and discuss their possible nature. We also detected a single source showing an extreme UV to optical color and corresponding to UV flare on a cool sdM subdwarf star. Finally, we discuss the possible populations of objects that may be revealed in a future larger-scale analysis of this kind.
\end{abstract}




\section{Introduction}
\label{sec:introduction}

The problem of parameterization of astronomical objects based on their photometry is
a topical issue. A great variety of photometric systems and
recently constructed large photometric surveys
as well as an emergence of dedicated VO tools for cross-matching their objects provide a unique possibility
to get multicolour photometric data for millions of objects.
This combined photometry can be used for relatively accurate determination of the parameters
of galaxies, stars, and the interstellar medium.
In particular, it was shown \citep{2018Galax...7....7M} that multicolour photometric data from large modern
surveys can be used for parameterization of stars closer than around 4.5 kpc and brighter than $g_{\rm SDSS}$ = 19.$^m$6,
including estimation of parallax and interstellar extinction value.

The objects detected in all the surveys under study
represent, naturally, the most favourable and convenient material for the research,
since the photometric data for them are most abundantly presented and cover the
electromagnetic spectrum from UV to IR.
However, objects that are found only in one of the surveys,
and absent in all others, should also attract close attention.

Here we present the pathfinder search for the UV objects from the GALEX survey
that have no apparent optical/IR counterparts, or have extreme UV to optical colors, based on an initial study of
a small fraction of the sky.
The paper is organized as follows. Section~\ref{sec:crossmatch} describes the selection of catalogues we use for the study and the initial quality cuts we applied, as well as the cross-matching procedure. In Section~\ref{sec:properties} the sample of UV-only objects, not matched with counterparts from optical catalogues, is presented. Next, in Section~\ref{sec:des} we build the UV-to-optical two-color diagram and investigate the objects there having extreme UV excess colors.
Section~\ref{sec:discussion} contains the discussion of the results and their possible physical implications, and Section~\ref{sec:conclusions} gives the conclusion of the study.

\section{Cross-matching of multi-wavelength surveys}
\label{sec:crossmatch}

\begin{figure*}
  \centering
  \centerline{
    \resizebox*{2\columnwidth}{!}{\includegraphics[angle=0]{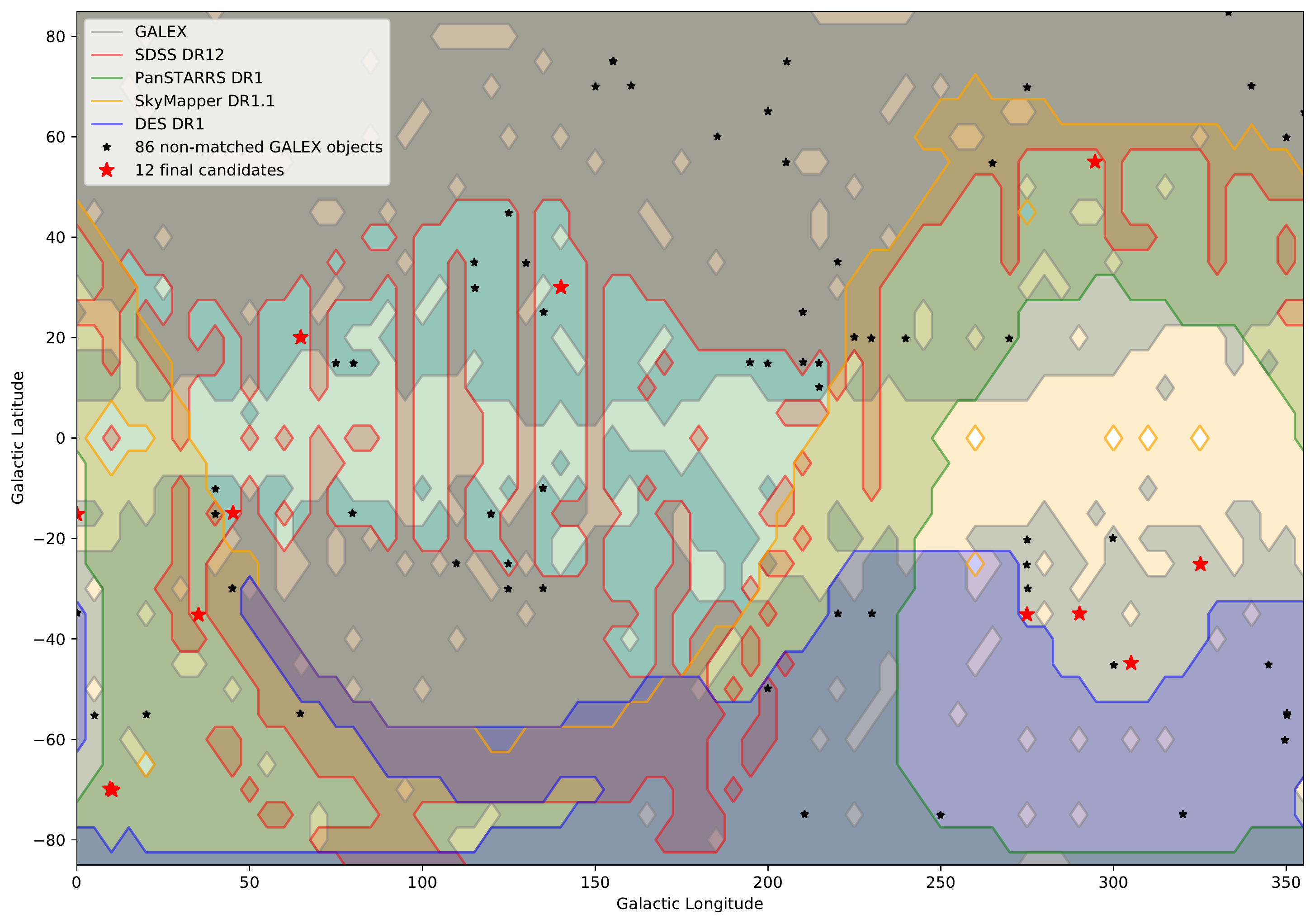}}
  }
  \caption{
    Sky footprint of GALEX and major optical sky surveys that do not provide all-sky coverage (SDSS DR12, PanSTARRS DR1, SkyMapper DR1.1, DES DR1). All other catalogs used in this study are covering the whole sky, and thus omitted from the plot.
    The footprint is plotted using small sample fields (0.25 deg radius) placed on a rectangular grid with 5 degree step in Galactic coordinates.
    Overplotted black stars represent the significant (S/N > 3 in both NUV and FUV bands) detections from GALEX catalogue not matched with optical catalogues, while red ones -- their final subset passing additional quality cuts as described in Section~\ref{sec:properties}.
  }
  \label{fig:spatial}
\end{figure*}

\begin{table*}
\caption{Summary of parameters of various sky surveys used in the present work. N$_{\rm fields}$ is the number of sky fields of our grid covered by the survey, with the next column showing the same information as a percentile of all sky fields. N$_{\rm stars}$ is the mean number of stars per field for a survey. P$_{\rm matched}$ is the fraction of GALEX objects (only the ones passing the initial criterion from Section~\ref{sec:crossmatch} are used here) matched to a given survey over all fields covered by it. P$_{\rm single}$ is the same fraction for GALEX objects matched with a given survey and not matched with all other surveys. Two final columns list the approximate depth of the surveys, along with the reference to the description paper.}
\label{tab:surveys}
\begin{tabular}{lrlccccr}
\hline
 & \multicolumn{2}{c}{N$_{\rm fields}$} & N$_{\rm stars}$ & P$_{\rm matched}$ & P$_{\rm single}$ & Depth & Reference \\
 & & & per field & \% & \% & & \\
\hline
GALEX (GUVcat\_AIS) & 1965 &  78\% & 574.1 &  &  & NUV = 21, FUV = 20 (AB, 5$\sigma$) & \citet{2017ApJS..230...24B} \\
SDSS DR12 & 1152 &  46\% & 9078.4 & 94.9 & 25.7 & $g = 23.2$, $r = 22.6$, $i = 21.9$ & \citet{2015ApJS..219...12A} \\
PanSTARRS DR1 & 2040 &  81\% & 8788.8 & 98.5 & 35.6 & $g = 23.2$, $r = 23.2$, $i = 23.1$ & \citet{2016arXiv161205560C} \\
SkyMapper DR1.1 & 1294 &  51\% & 1972.1 & 46.4 & 0.6 & $g = 21.7$, $r = 21.7$, $i=20.7$ & \citet{2018PASA...35...10W} \\
Gaia DR2 & 2520 & 100\% & 5502.1 & 36.0 & 11.9 & $G = 21$ & \citet{gaia} \\
WISE & 2520 & 100\% & 3427.1 & 85.5 & 5.0 & & \citet{2014yCat.2328....0C} \\
USNO-B1.0 & 2520 & 100\% & 4024.9 & 98.0 & 11.4 & $V=21$ & \citet{usno} \\
GSC2.3.2 & 2520 & 100\% & 3615.7 & 98.0 & 33.8 & $R_F=20.5$ & \citet{gsc} \\
DES DR1 & 495 &  20\% & 14477.5 & 97.7 & 68.0 & $g = 24.3$, $r = 24.1$, $i = 23.4$ & \citet{des} \\
\hline
 \end{tabular}
\end{table*}

In the process of studying interstellar extinction
\citep{Malkov2020},
we have cross-matched objects from various sky surveys in several selected sky areas and noticed the presence of a significant amount of objects present in GALEX catalogue only, without any counterparts in optical and infrared surveys. Therefore, we decided to make a dedicated study of these objects in order to assess their physical nature.

As the process of cross-matching several all-sky catalogues like GALEX is tricky and computationally intensive (see \citet{bianchi2020} for a successful example of such work, done with the aim opposite to the one in our analysis), we decided to start with the analysis of just a small subset of it in order to better understand the potential problems and formulate the exact criteria to be used for a later full-scale investigation.
Thus, we made a grid of 2520 sky fields, with 15 arcmin radius each, covering about 1\% of the whole sky and uniformly spaced in Galactic coordinates (5 degrees stepping in both directions).
For every field, we acquired from \textsc{VizieR} \citep{vizier} the lists of objects from the set of catalogues summarized in Table~\ref{tab:surveys}.
For that set, we selected the data products of major sky surveys (SDSS, PanSTARRS, SkyMapper, Gaia and WISE), all having uniform coverage of significant fractions of the sky, decent depth, uniform photometry in well-defined filters, and providing a representative multiwavelength coverage. In order to improve the characterization of fainter objects, we included two historical all-sky catalogues (USNO-B1.0 and GSC2.3.2) based on the digitization of photographic plates and providing good positional information. Also, in order to improve the coverage of the southern sky, we included the photometric catalogue of The Dark Energy Survey (DES DR1) covering approximately 5000 square degrees around South Galactic Pole.


Then, in every field, we cross-matched GALEX objects with all other catalogues using the pairwise distance threshold equal to the hypothenuse of positional accuracies of the two surveys. For the latter, we used the conservative value of 5$''$ (roughly two thirds of PSF, in contrast to 3$''$ used e.g. in \citet{bianchi2020}) for GALEX, 2$''$ for WISE, 0.1$''$ for Gaia DR2, and 1$''$ for all ground-based surveys, the same way as in \citet{2018Galax...7....7M} and \citet{Malkov2020}.
For an initial analysis presented here, we did not perform any comparison of object brightness in various catalogues, thus potentially excluding brighter UV objects where much fainter optical objects are present inside the matching circle.

A quick look study of the results revealed a great number of low-significance GALEX objects not having matched components in other sky surveys. Most of them represent, in our opinion, just a spurious detections\footnote{We leave the thorough investigation of this point and selection of a proper threshold that optimally separates spurious and real detections to a larger-scale follow-up work that will be based on the cross-match of a whole GALEX catalogue}. In order to reliably filter them out, we decided, for this initial analysis, to concentrate just on objects having both NUV and FUV detections\footnote{Despite worse FUV sensitivity of GALEX this criterion is reasonable as UV objects lacking bright optical counterparts should have quite steep spectra implying significant FUV components.} with S/N > 3, i.e. with e\_NUV$<$0.3 and e\_FUV$<$0.3. After applying such a cut-off, we decreased the total amount of GALEX objects in our sample from 1128082 to 38813, and  non-matched ones (we will call them ``UV-only'') -- from 222308 (19.7\% of all objects) to 86 (0.2\% of cut-off sample).

\section{Properties of UV-only objects}
\label{sec:properties}



\begin{figure}
  \centering
  \centerline{
    \resizebox*{1.0\columnwidth}{!}{\includegraphics[angle=0]{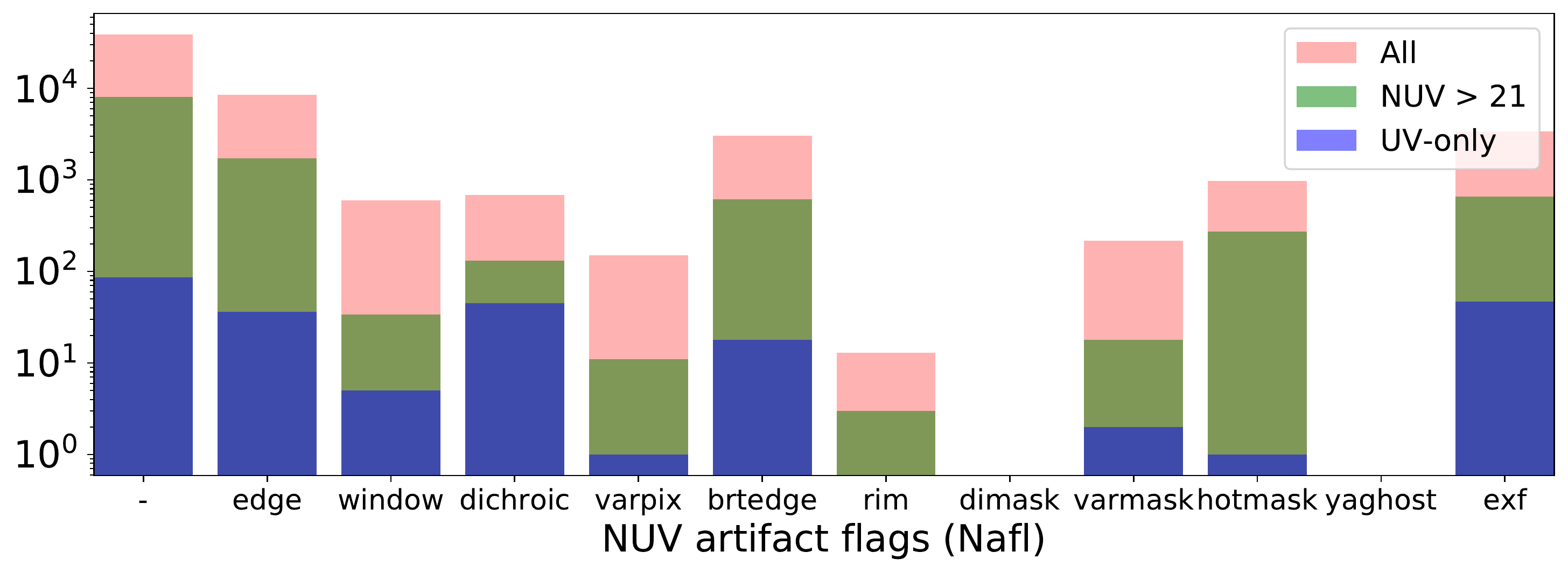}}
  }
  \caption{
  The distribution of objects by the NUV artefact flags (bits of Nafl, upper panel). 
  The last column also corresponds to the presence of extraction flags (Nexf).
    Red columns are all GALEX objects from the sky fields passing our initial selection criterion from Section~\ref{sec:crossmatch}, green are the subset of fainter (NUV > 21) objects, and blue ones -- UV-only objects.}
  \label{fig:NUV_flags}
\end{figure}

\begin{figure}
  \centering
  \centerline{
    \resizebox*{1.0\columnwidth}{!}{\includegraphics[angle=0]{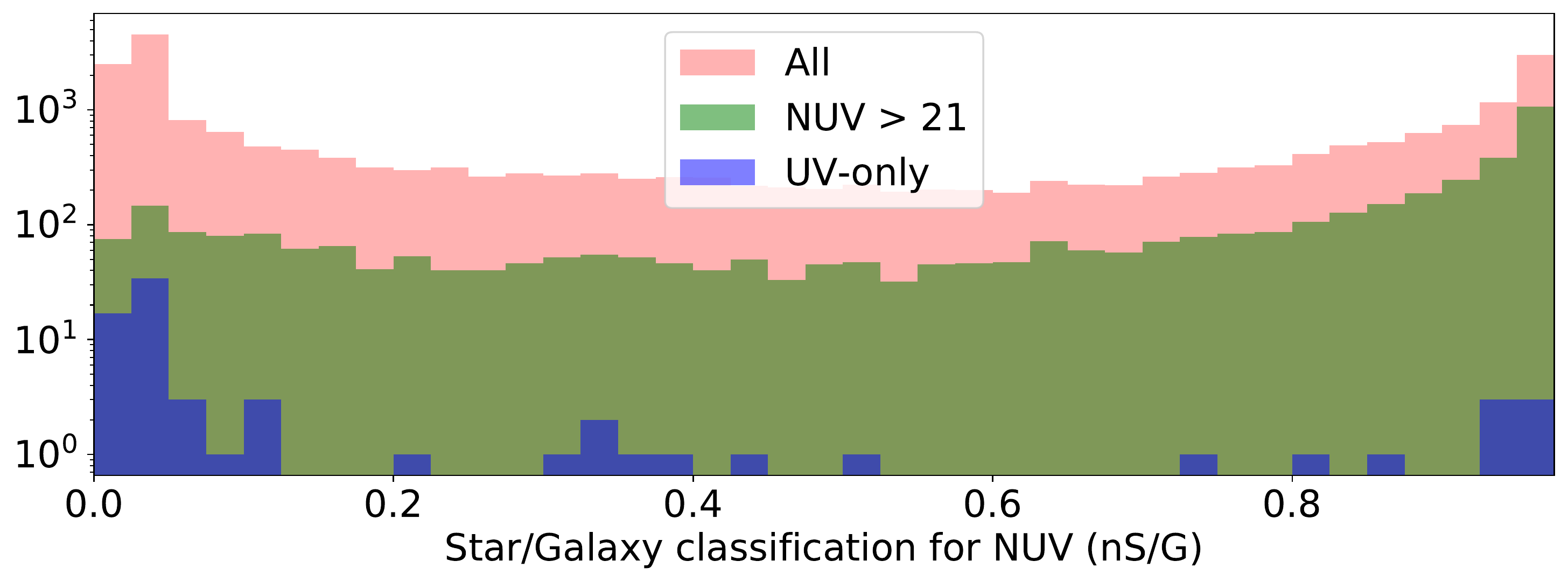}}
  }
  \caption{The distribution of objects by the NUV star/galaxy separation classification probability. The rightmost part corresponds to the highest probability of an object being point-like, while the leftmost -- to the ones with spatial extent clearly detectable in NUV. Red columns are all GALEX objects from the sky fields passing our initial selection criterion from Section~\ref{sec:crossmatch}, green are the subset of fainter (NUV > 21) objects, and blue ones -- UV-only objects.}
  \label{fig:NUV_star}
\end{figure}

\begin{figure*}
\centering
  \centerline{
    \resizebox*{1.0\columnwidth}{!}{\includegraphics[angle=0]{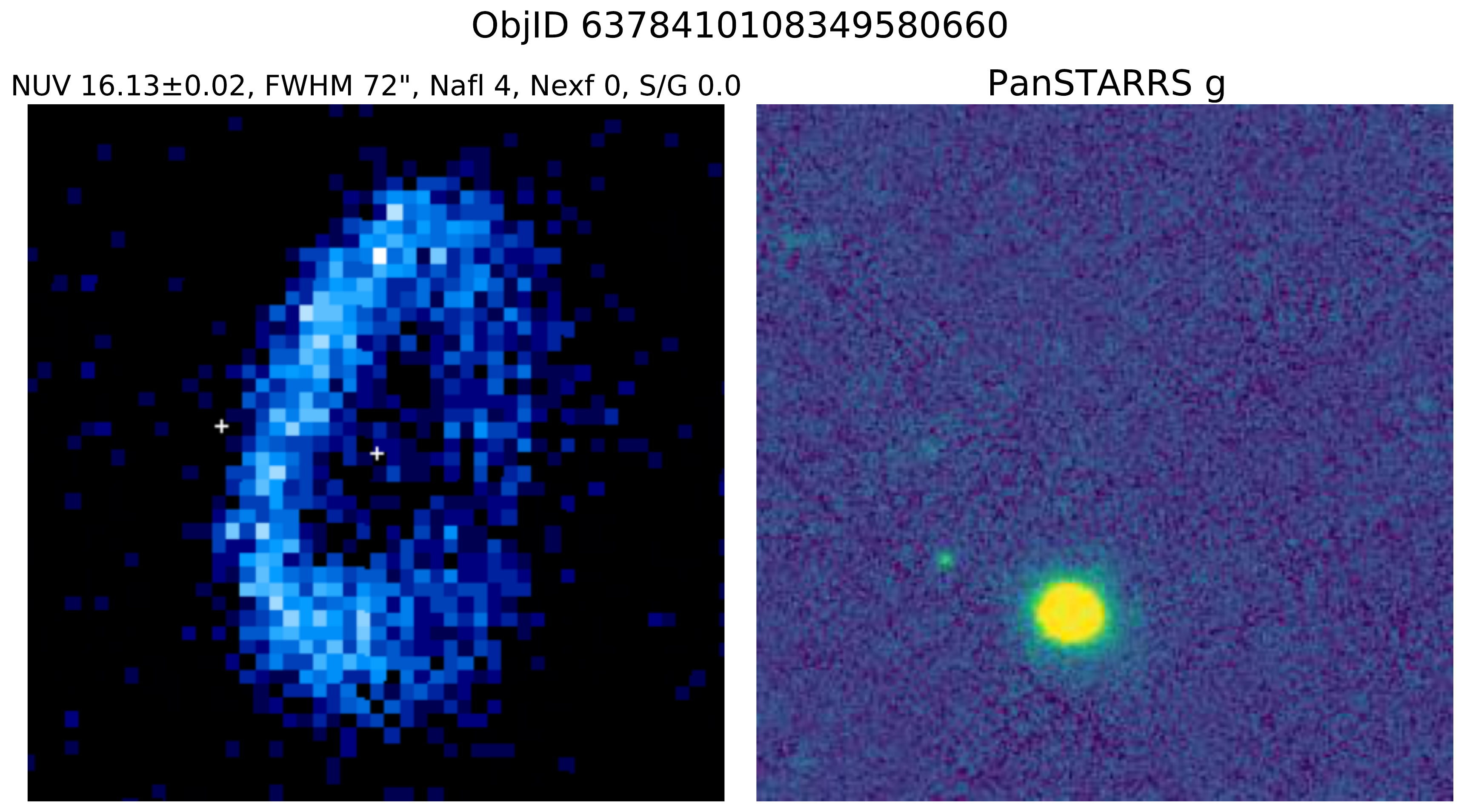}}
    \resizebox*{1.0\columnwidth}{!}{\includegraphics[angle=0]{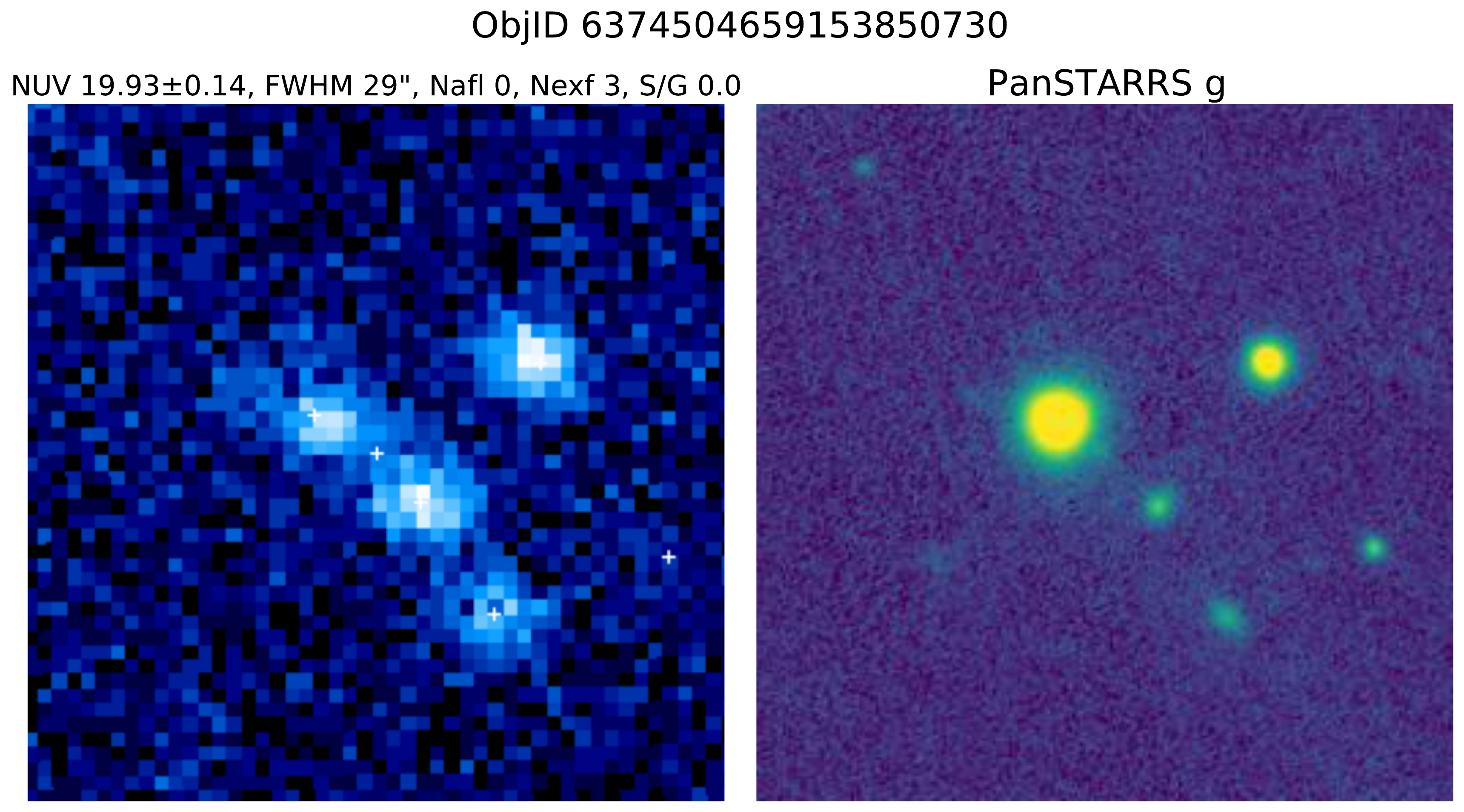}}
  }
  \caption{GALEX NUV and Pan-STARRS 72$''\times$72$''$ cutouts for some of the UV-only objects, corresponding to the dichroic reflection (left panel, 0x04 bit in Nafl marks this kind of artifacts) and deblending problems (right panel, 0x02 bit in Nexf marks deblended objects). White crosses denote the positions of GALEX catalogue sources. Both of these kinds of artifacts have low star/galaxy separation scores and correspond to the leftmost part of the histogram in Figure~\ref{fig:NUV_star}.
  The cutouts are produced using SkyView web service \citep{skyview}.}
\label{fig:artefacts}
\end{figure*}

\begin{figure}
\centering
  \centerline{
    \resizebox*{0.5\columnwidth}{!}{\includegraphics[angle=0]{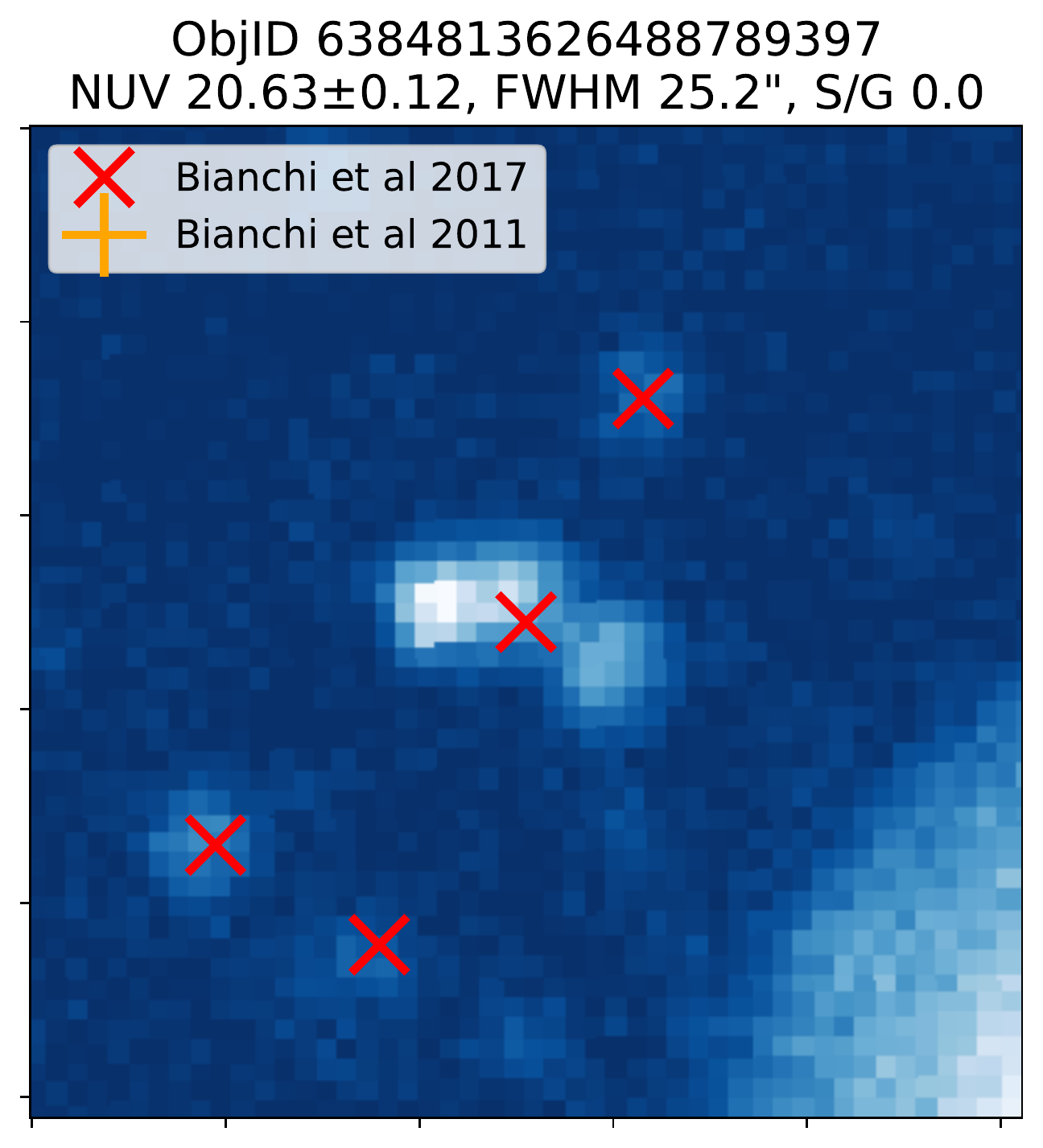}}
    \resizebox*{0.5\columnwidth}{!}{\includegraphics[angle=0]{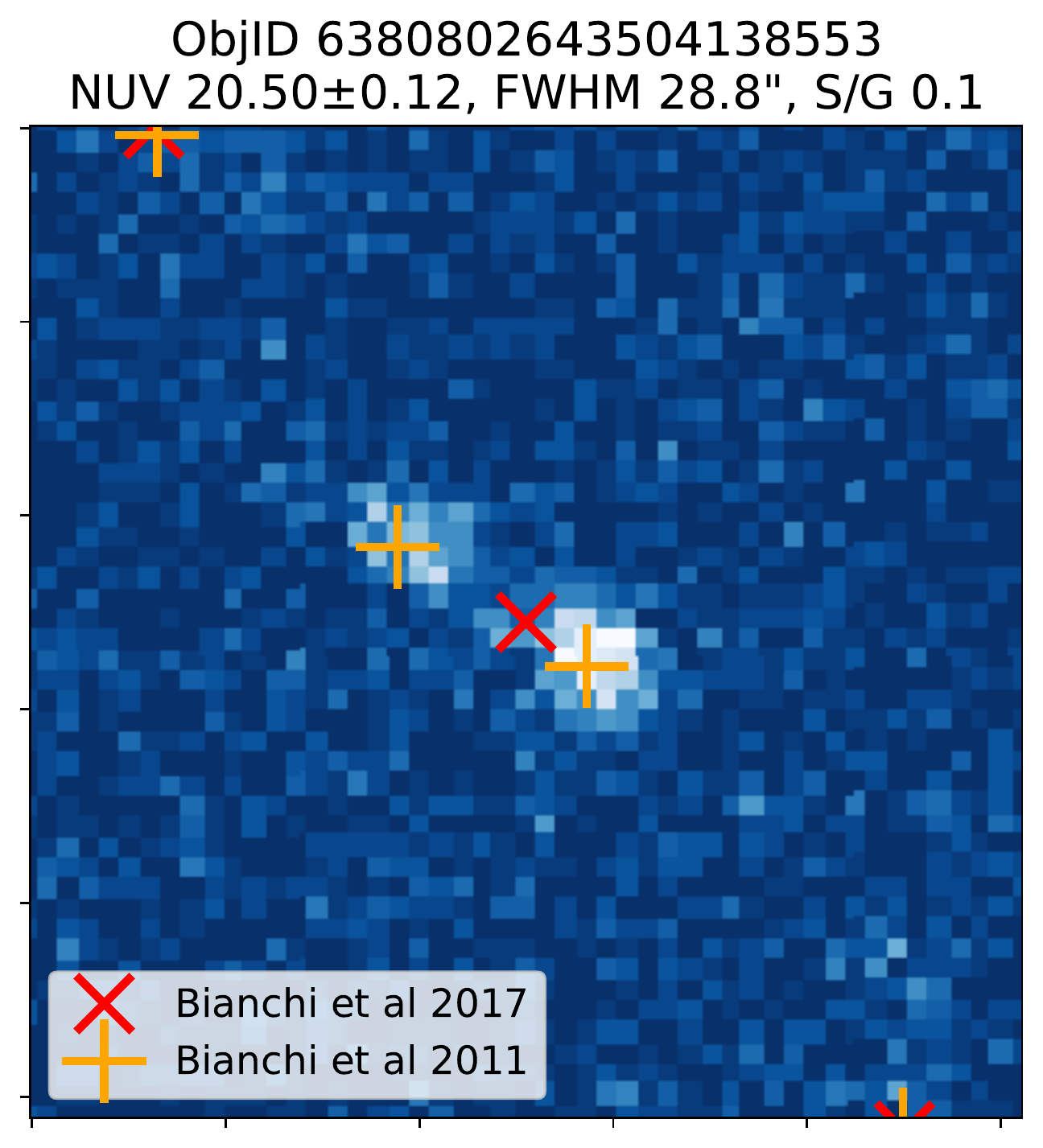}}
  }
  \caption{NUV images of two blended objects from GALEX catalogue. Overplotted are the positions of catalogue entries from \citet{2017ApJS..230...24B} and \citet{2011Ap&SS.335..161B}. On the left panel, the blend is not resolved in the newer catalogue version and not detected at all in the older one, while on the right panel -- again not resolved in the newer version but perfectly resolved in the older one. In both cases, the star/galaxy separation (nS/G) ratio is very small, clearly indicating an extended object.}
\label{fig:blends}
\end{figure}


Among the total of 38813 GALEX objects satisfying our quality criteria (see Section~\ref{sec:crossmatch}) in all fields where GALEX data are available, 86 (0.2\%) have no counterparts in other catalogues from our list, corresponding to other wavebands.
Hereafter, we will call these objects ``UV-only objects'' to distinguish them from the objects having such counterparts but showing extreme UV to optical colors (we will discuss them later in Section~\ref{sec:des}). 



To exclude various artifacts from the selected objects, we checked the quality indicators of the GALEX catalogue \citep{2017ApJS..230...24B} --
artifact flags (Nafl and Fafl) and extraction flags (Nexf and Fexf). Their distribution
is shown for both all sources from our original sample, a subset of fainter sources
(NUV > 21), and UV-only ones in Fig.~\ref{fig:NUV_flags}. There is no striking difference between these three classes of objects except for the slight increase in the frequency of dichroic related (bit 0x04 in Nafl) artifacts for UV-only objects. The latters are among two classes of artifact flags suggested to be excluded in Section 6.2 of \citet{2017ApJS..230...24B} (second one -- window edge reflections, corresponding to 0x02 bit in Nafl). Left panel of Figure~\ref{fig:artefacts} shows an example of a dichroic artifact among UV-only objects.

The presence of extraction flags also represents various kinds of potential problems, with the most prominent being the improper deblending, like the case in the right panel of Figure~\ref{fig:artefacts} where an extra object is erroneously detected between two blended components of a close group. On the other hand, the blend may be unresolved at all, being reported as a single extended object with no artifact or extraction flags set (see an example in Figure~\ref{fig:blends}).


GALEX photometric pipeline (based on SExtractor code by \citet{sextractor}) also reports the result of a star-galaxy (S/G) classification for all detection based on its fuzziness, corresponding to the probability of an object being point-source (S/G=1) or extended, galaxy-shaped (S/G=0). The distribution of these classifications for NUV band is shown in Fig.~\ref{fig:NUV_star}. Of all UV-only objects, only 21 (24\%) have a good probability to be point-source (nS/G$>$0.5).
The rest mostly represent artifacts like the ones discussed in the previous paragraph, or truly spatially extended objects like galaxies. 
Both these classes are unsuitable for our current work, as we are cross-matching with primarily point-source catalogues. Thus, we will concentrate solely on the analysis of point-source objects with nS/G$>$0.5. To that, we add the requirement for the absence of NUV dichroic reflection flags (Nafl\&0x04=0), and the absence of NUV and FUV extraction flags (Nexf=0 and Fexf=0). After such filtering, only 12 objects remain in the final sample, which is presented in Table~\ref{tab:candidates}.

Among them, none is located inside SDSS DR12 or DES DR1 footprints, while 7 are covered by PanSTARRS DR1 and 9 -- by SkyMapper DR1.1. That suggests that the main reason for those objects being undetected may be their faintness -- they are below the detection threshold of the two latter catalogues (especially SkyMapper which is the least sensitive among them), and at the same time are above the limit of SDSS and especially DES.

Deep co-added PanSTARRS DR1 images are available for 5 of 7 candidates covered by PanSTARRS survey footprint (two more are located below the declination of -30 degrees, where co-adds are not created despite catalogue data still being available). Visual inspection of four of them (\#4, \#8, \#9 and \#10) reveals faint blueish optical uncatalogued sources at the positions of candidates. We performed simple aperture photometry of their fields, calibrated the zero point using PanSTARRS DR1 catalogue entries for several tens of neighbour stars, and estimated the $g$, $r$ and $i$ magnitudes of these objects. The results listed in the Comments below Table~\ref{tab:candidates} are marginally brighter than the formal depth of PanSTARRS DR1 catalogue listed in Table~\ref{tab:surveys}, suggesting that there are some additional selection effects preventing their detection by the official pipeline that explains why these candidates are not matched.

One more candidate, \#5, is located on the spiral arm of NGC~4504. Visual inspection of its GALEX data reveals no point source at the object location, however, with a nearby apparent uncatalogued object consistent with $g$=21.5 PanSTARRS stars at a 8$''$ distance from the candidate position. We thus suggest that this candidate is a result of an erroneous position measurement, probably due to rapidly varying background due to the nearby galaxy.

Finally, for two candidates in PanSTARRS footprint not covered by deep co-added images, there are faint counterparts in an infrared VISTA Kilo-degree INfrared Galaxy (VIKING, \citet{viking}) and optical OmegaCAM Kilo-Degree Survey (KiDS, \citet{kids}) surveys, with one (\#6) being point source, while the other (\#12) clearly showing an extended galaxy-like shape and an overall SED shape typical for active galaxies.

Two more objects (\#1 and \#2) have infrared point-like counterparts in VISTA Magellanic Survey (VMC) DR4 \citep{vmc} survey, with the SED marginally consistent with being a hot massive star.

Finally, three candidates (\#3, \#7 and \#11) are completely undetected on longer wavelengths, being outside of the footprints of deep smaller area southern surveys like VIKING or KiDS.


\begin{table*}
\caption{UV-only objects that do not have optical counterparts in Gaia DR2, PanSTARRS DR1, SDSS DR12, SkyMapper DR1.1 and WISE catalogues. ObjID is an object identifier according to \citet{2017ApJS..230...24B}, NUV and FUV are catalogue magnitudes in corresponding energy bands, while Nr and Fr are corresponding source FWHMs in the same bands. The last column refers to the comments given below the table that list the results of additional investigation of these objects.}
\label{tab:candidates}
\begin{tabular}{llccccccr}
  \hline
  & ObjID & RA & Dec & NUV & FUV & Nr & Fr & Comments \\
&& hours & degrees & mag & mag & arcsec & arcsec & \\
\hline
\#1 & 6385728416941869657 & 05 09 37.3843 &  -64 51 26.1072 &  20.29 $\pm$ 0.15 & 19.98 $\pm$ 0.18 & 10.8 & 7.2 & 1 \\
\#2 & 6384919152761504385 & 00 31 04.1513 &  -72 14 59.604 &  20.57 $\pm$ 0.17 & 20.76 $\pm$ 0.20 & 18.0 & 18.0 & 2 \\
\#3 & 6383934016112821437 & 18 49 44.8061 &  -35 52 48.3816 &  20.79 $\pm$ 0.22 & 20.28 $\pm$ 0.20 & 14.4 & 7.2 & \\
\#4 & 6375947121772727620 & 18 28 47.7022 &  +36 41 12.6384 &  20.98 $\pm$ 0.23 & 20.82 $\pm$ 0.26 & 18.0 & 10.8 & 3 \\
\#5 & 6382561811642715789 & 12 32 21.2878 &  -07 31 39.4356 &  21.13 $\pm$ 0.23 & 21.38 $\pm$ 0.28 & 21.6 & 14.4 & 4 \\
\#6 & 6380732277981190131 & 23 23 22.6891 &  -33 03 40.4208 &  21.58 $\pm$ 0.17 & 21.23 $\pm$ 0.17 & 18.0 & 7.2 & 5 \\
\#7 & 6387628406320666178 & 18 48 42.1805 &  -69 56 56.8356 &  21.82 $\pm$ 0.30 & 21.13 $\pm$ 0.24 & 14.4 & 7.2 & \\
\#8 & 6379395228892139165 & 20 06 51.4646 &  +03 46 09.9516 &  21.96 $\pm$ 0.26 & 21.08 $\pm$ 0.25 & 10.8 & 14.4 & 6 \\
\#9 & 6379641490505734004 & 21 02 53.4823 &  -13 32 22.5888 &  22.03 $\pm$ 0.24 & 21.16 $\pm$ 0.19 & 14.4 & 7.2 & 7 \\
\#10 & 6373449093631448012 & 07 51 24.2273 &  +74 40 12.7524 &  22.14 $\pm$ 0.30 & 21.92 $\pm$ 0.28 & 10.8 & 10.8 & 8 \\
\#11 & 6385059894544830558 & 04 16 27.6175 &  -76 34 41.5488 &  22.25 $\pm$ 0.27 & 21.12 $\pm$ 0.30 & 7.2 & 10.8 &  \\
\#12 & 6380732277981188022 & 23 22 21.3468 &  -33 19 27.174 &  22.72 $\pm$ 0.29 & 22.21 $\pm$ 0.27 & 7.2 & 14.4 & 9 \\
\hline
 \end{tabular}

 {\raggedright
    1. This object has a pair of J=21.9 and J=22.7 mag point-like counterparts within 3$''$ in VISTA Magellanic Survey (VMC) DR4 \citep{vmc}. \\
    2. There is also a J=21.8 point-like source at 4$''$ in VMC DR4 \citep{vmc} \\
    3. This object has a faint uncatalogued point-like counterpart visible in PanSTARRS g, r and i band images at a 2$''$ distance, with rough magnitude estimates of $g$=22.45$\pm$0.11, $r$=22.73$\pm$0.18 and $i$=23.5$\pm$0.3. \\
    4. The object is located on the outer spiral wing of NGC 4504. This object has the lowest (nS/G=0.75) star/galaxy rating among the sample. Visual inspection of the GALEX NUV cutout does not show any source at the position, but shows an uncatalogued point source at 8$''$ corresponding to $g$=22.5 PanSTARRS star, so we suggest this source to have a wrong position in the catalogue. \\
    5. This object has $g$=21.78$\pm$0.02, $r$=22.3$\pm$0.03 and $i$=22.46$\pm$0.12 stellar counterpart in OmegaCAM Kilo-Degree Survey (KiDS) DR3 survey \citep{kids}, as well as Z=22.4 mag counterpart in VISTA Kilo-degree INfrared Galaxy (VIKING) DR2 survey \citep{viking} \\
    6. This object has a faint uncatalogued point-like counterpart visible in PanSTARRS g, r and i band images, with  rough magnitude estimates of $g$=22.7$\pm$0.08, $r$=22.8$\pm$0.2 and $i$=23.4$\pm$0.4 \\
    7. This object has a faint uncatalogued point-like counterpart visible in PanSTARRS g, r and i band images, with  rough magnitude estimates of $g$=22.3$\pm$0.1, $r$=22.8$\pm$0.2 and $i$=23.9$\pm$0.4 \\
    8. There is a faint uncatalogued point-like counterpart visible in PanSTARRS g and r band images at a 2$''$ distance, with  rough magnitude estimates of $g$=22.6$\pm$0.1 and $r$=22.8$\pm$0.2 \\
    9. This object has extended $g$=21.76$\pm$0.03, $r$=21.14$\pm$0.02 and $i$=20.87$\pm$0.05 counterpart in OmegaCAM Kilo-Degree Survey (KiDS) DR3 survey \citep{kids}, and J=19.5 mag counterpart in VISTA Kilo-degree INfrared Galaxy (VIKING) DR2 survey \citep{viking}, with extended emission also clearly visible in the corresponding cutout. \\
  \par}
\end{table*}

\section{Objects with extreme UV-optical colors}
\label{sec:des}

\begin{figure*}
\centering
  \centerline{
    \resizebox*{2.0\columnwidth}{!}{\includegraphics[angle=0]{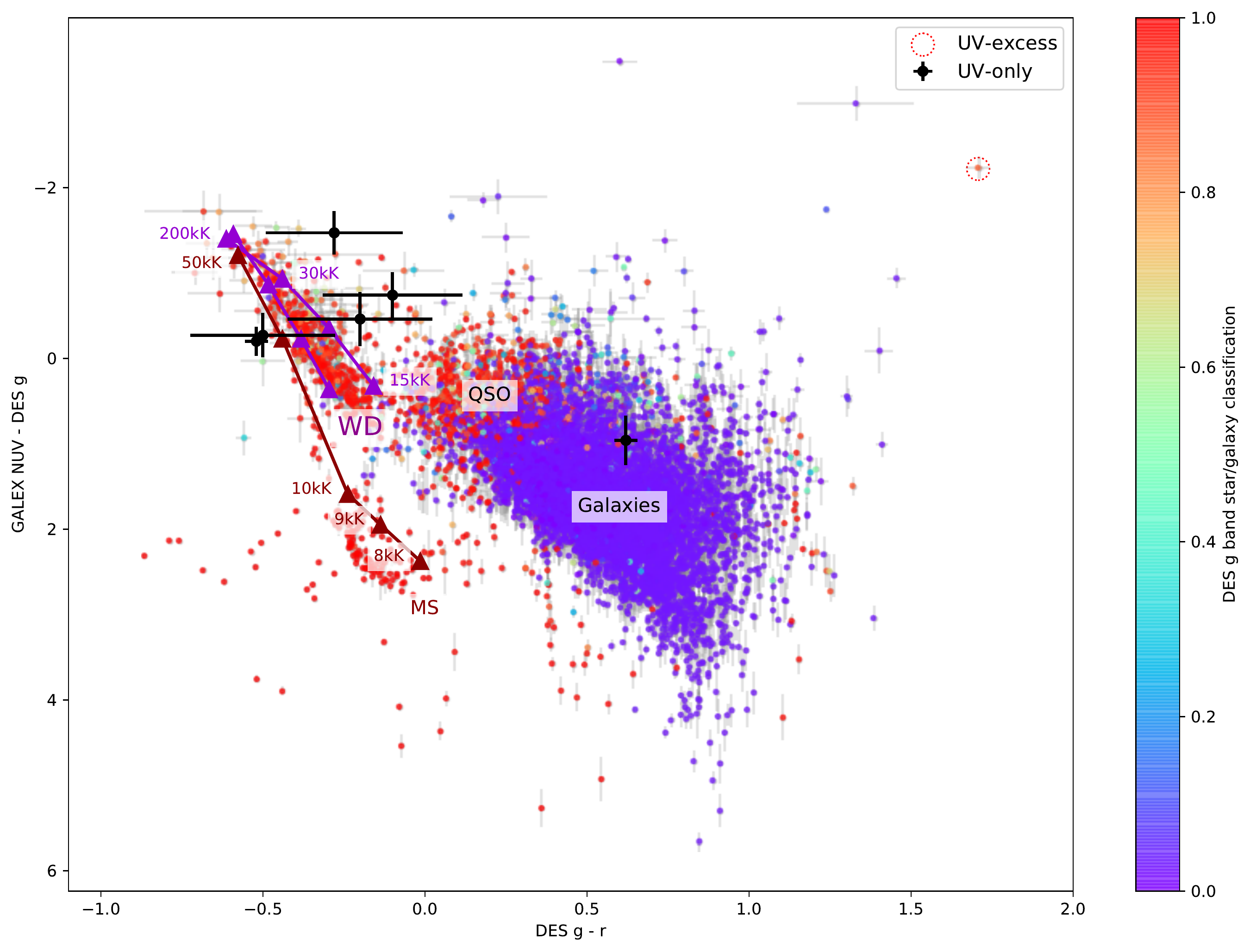}}
  }
  \caption{UV-to-optical two-color diagram for the objects cross-matched between GALEX and DES DR1 catalogues. Point color corresponds to the DES g band star/galaxy classification rating, with blue corresponding to extended sources, and red -- point-like ones. Black circles with error bars correspond to the objects from Table~\ref{tab:candidates} where g and r band photometry is available (see comments there). Red dotted circle denotes the only point- source object with (NUV-$g$)<-2 (see text). Dark violet triangles mark the sequence of white dwarfs of various temperatures and metallicities, while dark red ones -- the sequence of hottest main sequence stars, both taken from \citet{bianchi2020}. Also, the loci of QSOs and galaxies, also taken from the same work, are marked with the text labels.}
\label{fig:des_gr}
\end{figure*}

\begin{table*}
\caption{The single object with extreme UV to optical colors from the cross-matching of GALEX catalogue with DES DR1. All columns are taken from GALEX data except for S/G$g$ star/galaxy classification ratio and $g$, $r$ and $i$ magnitudes whose are taken from DES DR1 data.}
\label{tab:des}
\begin{tabular}{llccccccccc}
  \hline
  & ObjID & RA & Dec & NUV & FUV & S/G$g$ & $g$ & $r$ & $i$ \\
  & & hours & degrees & mag & mag & & mag & mag & mag \\
  \hline
\#1 & 6384602553542247330 & 02 09 30.26 & -53 35 12.9 & 20.25$\pm$0.13 & 20.65$\pm$0.21 & 0.875 & 22.47$\pm$0.03 & 20.76$\pm$0.01 & 18.89$\pm$0.00 \\
  \hline
 \end{tabular}
\end{table*}

\begin{figure}
\centering
  \centerline{
    \resizebox*{1.0\columnwidth}{!}{\includegraphics[angle=0]{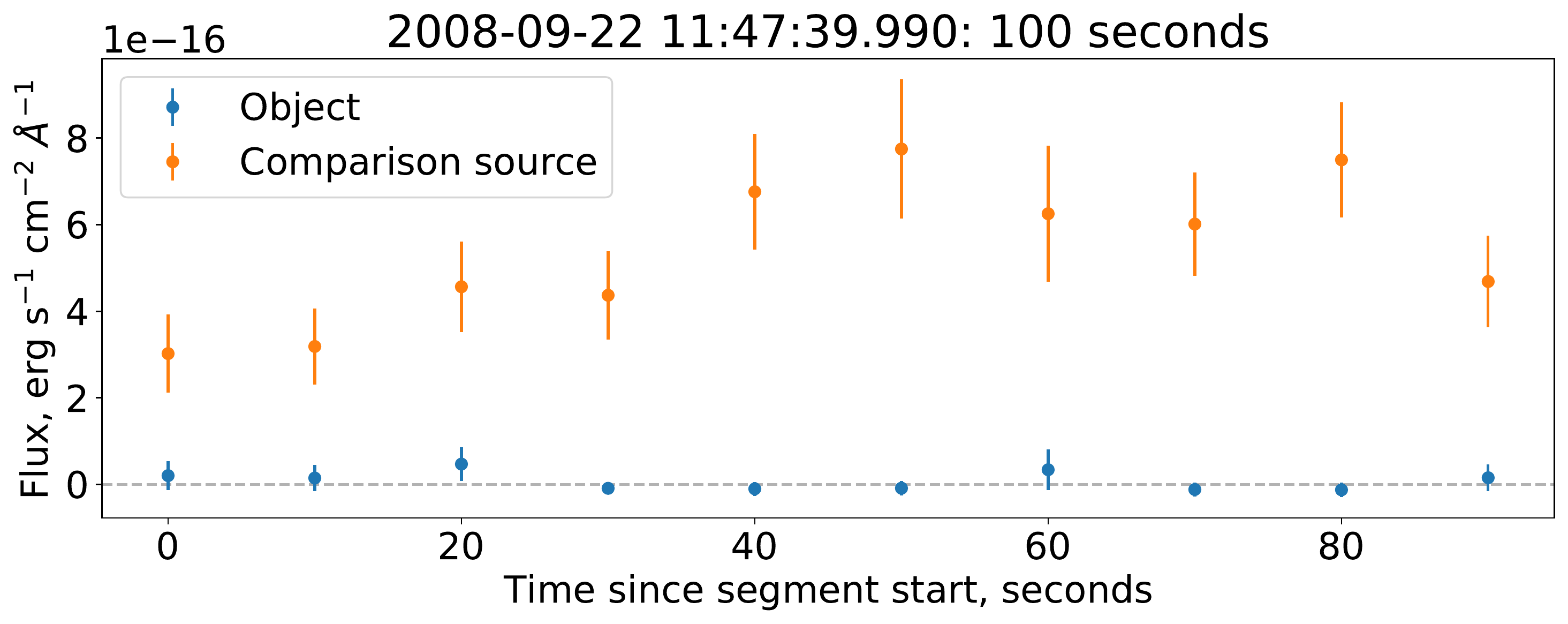}}
  }
  \centerline{
    \resizebox*{1.0\columnwidth}{!}{\includegraphics[angle=0]{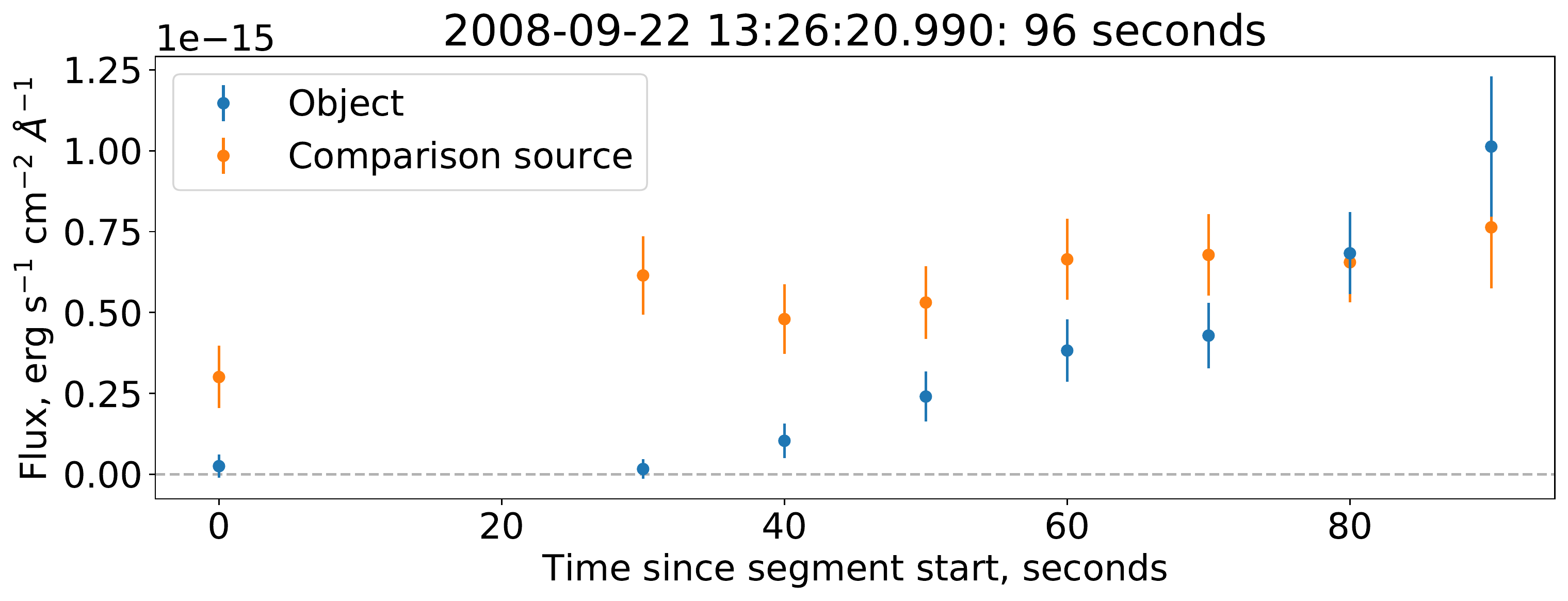}}
  }
  \centerline{
    \resizebox*{1.0\columnwidth}{!}{\includegraphics[angle=0]{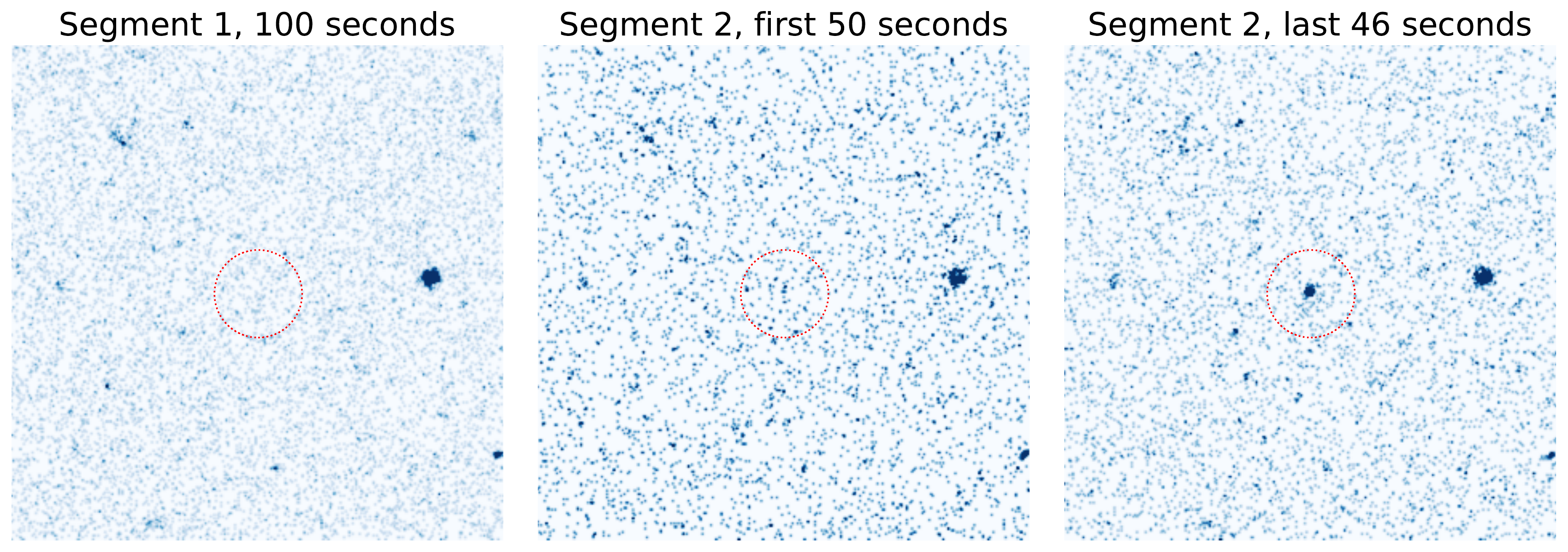}}
  }
  \caption{NUV light curves (two upper panels) of an object with extreme UV-optical color from Table~\ref{tab:des} and a nearby supposedly stable UV source (NUV=18.27$\pm$0.05). The data are acquired and calibrated using \textit{gPhoton} software package by \citet{gphoton}. For that, we chose the 5$''$ aperture radius, background annulus spanning between radii of 10$''$ and 60$''$, and a time step of 10 seconds (approximately 1/10 of the data segment length). We then excluded from the light curves the points marked as unreliable or problematic by \textit{gPhoton}.
  The object position was observed twice during the GALEX mission. During the first data segment, its NUV emission is not detectable, while during the second one an apparent flare onset is clearly visible in its light curve. Visual inspection of the time-resolved images (lower panel shows low temporal resolution coadds of first and second data segments, but we also checked higher resolution ones visually) built also with \textit{gPhoton} confirms that it is indeed a flash of the positionally stable source, and not a moving object or some imaging artifact polluting the aperture. The brighter source to the right is the one used for comparison in the light curves.}
\label{fig:sdm_lc}
\end{figure}

Among 12 candidate objects selected in Section~\ref{sec:properties} as ``UV-only'', i.e. not having counterparts in major optical catalogues listed in Table~\ref{tab:surveys}, 8 are actually showing the counterparts at longer wavelengths after a more detailed analysis (with one more most probably being the result of position determination error). Six of them have optical brightness measurements in $g$, and $r$ filters with sufficient accuracy.  Therefore, in order to better assess the parameters of these objects, we decided to study their locations on UV-to-optical two-color diagram.

To do so, we selected the deepest catalogue in our sample, DES DR1 \citep{des}, and again cross-matched it with GALEX, this time using a match radius of 3$''$, which is an optimal radius according to \citet{bianchi2020}. In order to exclude artificially large UV-to-optical colors, we selected all ambitious (non-unique) matches and kept only the ones among them with the brightest optical components (this way we may bias the colors towards UV deficiency, which is acceptable if we are looking primarily for objects with UV excess).
Moreover, we again excluded the GALEX objects with S/N<3, i.e. e\_NUV>0.3 or e\_FUV>0.3, and the objects with dichroic artifact flags or extraction flags. We also excluded DES DR1 objects with $g$ or $r$ band isophotal flags (gIsoFl and rIsoFl) set, and kept only the ones with recommended values of extraction flags (gFlag$<$4 and rFlag$<$4).

Figure~\ref{fig:des_gr} shows the distribution of the resulting matches in a two-color DES ($g$-$r$) vs (GALEX NUV - DES $g$) diagram, with color-coded DES g band star/galaxy classification rating clearly separating the loci of stellar (point sources) from extragalactic (extended) objects, with also a well-defined cloud of point-like QSOs in the middle.  The plot is mostly analogous to Figures 4 and 5 of \citet{bianchi2020} and has the same overall layout. The positions of 6 objects from Table~\ref{tab:candidates} where $g$ and $r$ magnitudes are available, are shown with black circles with corresponding error bars. The only one (\#12) having extended optical emission falls into the extragalactic region, while five others are consistent with the locus of the hottest stars. Unfortunately, the accuracy of the measurements does not allow to pinpoint them to either main sequence or hot white dwarfs tracks.

Nevertheless, the colors of these objects are quite typical and do not represent any extreme population. Thus, we decided to investigate the outliers in the two-color diagram of Figure~\ref{fig:des_gr} corresponding to the extreme UV excess (we will call them ``UV-excess'' objects). As a numerical criterion for the latter, we chose (NUV-$g$)$<$-2, which exceeds, within typical error bars, the values for the hottest main sequence stars and white dwarfs\footnote{The sequences of white dwarfs and main sequence stars of various temperatures are shown in Figure~\ref{fig:des_gr} after \citet{bianchi2020}. These locations are for unreddened case -- however, the reddening shifts the points there towards bottom-right, and thus can\'t move ``normal'' object above the limit for hottest stars}, as well as the locus of extragalactic sources. We also require the objects to be point-like by having S/Gg$>$0.5.

After applying these criteria, we got only one source whose parameters are listed in Table~\ref{tab:des}. It is GALEX~6384602553542247330 positionally coincident with DES~J020930.29-533512.6, with (NUV-$g$)=-2.22$\pm$0.13, i.e. marginally consistent with our color criterion. The object also has marginal red colors, ($g$-$r$)=1.70$\pm$0.03, ($r$-$i$)=1.87$\pm$0.01 and ($r$-$z$)=2.69$\pm$0.01, placing it well outside the locus of hot stars and suggesting that it is a cool sdM subdwarf \citep{savcheva2014}. The Gaia EDR3 distance estimate of r=337~pc \citep{bailerjones2021}, giving absolute magnitude M$_{r}$=13.12 is also consistent with this interpretation.

While strong UV emission is not expected from such cool stars, magnetic activity is quite prominent in them \citep{savcheva2014}. Thus we decided to investigate the UV emission from this object in more detail. For it, we constructed the light curves and time-resolved images of the object and a nearby brighter UV source using \textit{gPhoton} software package \citep{gphoton} over all two of the time segments (both approximately 100 seconds long) when the position was observed by GALEX. The light curves are shown in Figure~\ref{fig:sdm_lc}, and it is clearly seen there that the source is essentially invisible (and its flux is consistent with the background level) in the first data segment and during the first half of the second one. However, since approximately 40 seconds into the second segment, the brightness of the source starts to rapidly rise, becoming even brighter than the comparison object (which has NUV=18.27$\pm$0.05) during the last 10 seconds.
Visual inspection of the time-resolved images (lower panel of Figure~\ref{fig:sdm_lc}) also confirms that the event is indeed a flare from a point source, and not some moving object or imaging artifact passing the aperture. Thus we may conclude that it is indeed a stellar flare, with a timescale and amplitude not unusual for typical flares observed by GALEX \citep{galex_flares}.

\section{Discussion}
\label{sec:discussion}

Our initial search did not reveal any high significance (S/N $\gg$ 5) detection of GALEX objects not matched with optical catalogues -- UV-only objects. However, several lower significance ones (with S/N around 3 to 5, but having simultaneous detection in both NUV and FUV bands, not marked as imaging artifacts or blends, and thus most probably corresponding to actual astrophysical objects) UV-only objects (UVLO candidates) are detected (see Table~\ref{tab:candidates}).
They are all located outside of sky regions covered by the deepest surveys from our sample -- SDSS DR12 and DES DR1. However, two of them have faint catalogued counterparts in an infrared VISTA Kilo-degree INfrared Galaxy (VIKING, \citet{viking}) and OmegaCAM Kilo-Degree Survey (KiDS, \citet{kids}) surveys, while for 4 more we were able to locate and measure faint uncatalogued counterparts in PanSTARRS DR1 deep co-added images. Among these six objects with measured optical colors, five are spanning the locus of hot massive stars in Figure~\ref{fig:des_gr}, while one -- the locus of galaxies. Two more objects also have infrared point-like counterparts in VISTA Magellanic Survey (VMC) DR4 \citep{vmc} survey, with the SED also marginally consistent with being a hot massive star. Of the remaining four, one is most probably the result of improper position determination, while final three (\#3, \#7 and \#11) are completely undetected on longer wavelengths, being outside of the footprints of deeper southern sky surveys like VIKING and KiDS.

Our search for GALEX objects what do have optical counterparts but display an unusual -- extreme -- UV to optical color in the footprint of DES DR1 survey, the deepest among the ones considered in this work, has revealed a single source (see Table~\ref{tab:des}) with the (NUV-$g$) color exceeding the one for the hottest main sequence stars and white dwarfs. Its detailed study revealed that the object is actually a cool sdM subdwarf, with UV emission apparent during the flare-like event, which we attribute to its magnetic activity.

Thus, only one of twelve UV-only objects we selected shows an extended emission at longer wavelengths and an AGN-like SED, while five more occupy the locus of hot stars on a color-color diagram (like the ones shown in Figures~4 and 5 of \citet{bianchi2020}). Indeed, one of the most obvious candidates for ultraviolet luminous objects (UVLO) is hot massive stars. However, even the hottest stars would have (GALEX NUV - SDSS $g$) color of -1.5 or smaller \citep{bianchi2011}, thus they should also be sufficiently bright in the optical range to be detectable by the modern sky surveys like SDSS or KiDS, while occasionally invisible in the wider field but less sensitive ones like PanSTARRS 3Pi Survey and especially SkyMapper.

The hottest massive stars on the main sequence have $M_{V}=-6.35$ and thus Galactic ones should be extremely bright unless highly reddened, therefore we may expect only the ones in nearby galaxies to pass our criteria of being non-detectable on longer wavelengths.
On the other hand, hot white dwarfs may be as faint as $M_{V}$=9..12, and their population in the Solar vicinity (tens of pc up to kpc) may constitute a large fraction of the UVLO.
On the more exotic side, strong UV excess is observed in a recent detection of a massive super-Chandrasekhar merger product of binary white dwarf system \citep{gvaramadze2019}.

Another possible candidate for the role of UVLO could be
isolated (single or components of wide binaries) old neutron stars, slowly accreting interstellar matter.
This type of objects was proposed by \citet{1997MNRAS.287..293M} as candidates
to the sources detected in the extreme ultraviolet by the ROSAT/WFC
and EUVE all-sky surveys and unidentified with any optical counterpart.
Such neutron stars are expected to be abundant in the Galaxy,
to concentrate towards the Galactic plane, and have not been unambiguously detected yet.
Even more abundant, and also still undetected, is the Galactic population of isolated stellar-mass black holes produced as a result of stellar evolution. They may produce faint synchrotron UV and X-ray emission due to the spherical accretion of magnetized interstellar matter \citep{beskin2005} and thus also contribute to the UVLO population.



Finally, we note that our analysis should be able to reveal, apart from the objects with strong UV excesses in the spectra, also the transient ones -- the objects that were either only visible during the GALEX observations of their position (and not during all other surveys), or the moving ones corresponding to Solar system bodies. \citet{asteroids} found 1342 detections of 405 asteroids appearing in GALEX images. Several of them fall within our sky fields, and 5 have spatially coincident entries in the revised version of GALEX catalogue \citep{2017ApJS..230...24B} we are using. However, none of them has FUV detections (which is consistent with \citet{asteroids}), thus failing our initial quality check. It stresses the necessity of a more sophisticated treatment of spurious events necessary for a full-scale analysis that we are planning as a continuation of current work.

Transient events detectable by GALEX naturally include flares on magnetically active stars. The search for such events has been performed e.g. by \citet{galex_flares} who have been able to detect 1904 short-duration flares on 1021 stars, with amplitudes of flux enhancement reaching 1700 times above quiescent levels. Such events should also appear as either UV-only objects (for distant low-luminosity stars in the Galaxy below the detection limit of the deepest optical sky surveys), or sources with extreme UV to optical color (for closer ones, like the flare on a cool sdM subdwarf we detected in Section~\ref{sec:des}).

Other transient sources, which may be detected in the ultraviolet spectral region, are UV-outbreaks prior to SNe.
The radius and surface composition of an exploding massive star,
as well as the explosion energy per unit mass, can be measured using
early UV observations of core-collapse SNe.
A theoretical framework to predict the number of early UV SN detection
in GALEX and planned ULTRASAT~\citep{2014AJ....147...79S}
surveys was developed by~\citet{2016ApJ...820...57G}, and the
comparison of observations with calculated rates shows a good
agreement.
Also, it was found by~\citet{2016ApJ...820...57G} that seven SNe were
clearly detected in the GALEX NUV data.
Other astrophysical sources potentially yielding a transient UV signal
are gamma-ray burst early afterglows, tidal disruption events, and AGN flares.



\section{Conclusions}
\label{sec:conclusions}

We performed a pathfinder study aimed at the characterization of the objects apparent in UV bands but lacking counterparts at longer wavelengths, or displaying extreme UV to optical colors --  ``ultraviolet luminous objects'' (UVLO). To do so, we cross-matched the catalogue of GALEX sources with several other modern large-scale sky surveys in a number of fields covering about 1\% of the sky. Even with quite restrictive quality cuts (which undoubtedly should be relaxed for the follow-up larger scale analysis) we have been able to uncover several faint UVLO candidates, and a single transient event corresponding to the flare on a cool sdM subdwarf, what testifies the approach we use.




Our initial analysis demonstrated the importance of such a study, and we plan to continue it using larger-scale cross-matching of the whole GALEX catalogue with other all-sky surveys, combined with smarter selection criteria and methods for filtering out spurious events
in order to reliably uncover the UVLO population, detect individual UVLO objects and perform their detailed investigation.


\section*{Acknowledgements}

We thank James Wicker for valuable comments.
S.K. was supported by European Structural and Investment Fund and the Czech Ministry of Education, Youth and Sports (Project CoGraDS -- CZ.02.1.01/0.0/0.0/15 003/0000437). The work is partially performed according to the Russian Government Program of Competitive Growth of Kazan Federal University.
O. M. thanks the CAS President's International Fellowship Initiative (PIFI).
Part of this work was supported by the National Natural Science Foundation of China
under grant Nos. 11988101, 11890694 and the Chinese-Russian NSFC-RFBR project number 20-52-53009.
and use of the VizieR catalogue access tool, CDS, Strasbourg, France.


\section*{Data availability}

The data underlying this article will be shared on reasonable request to the corresponding author.



\bibliographystyle{mnras}
\bibliography{uvlo}




%
%


\bsp	
\label{lastpage}
\end{document}